\documentclass[useAMS,usenatbib]{mn2e}
\usepackage{graphicx,amssymb, txfonts,pstricks}
\usepackage[dvips]{epsfig}
\usepackage{subfig}
\usepackage{float}

\def\feii{[Fe\,{\sc ii}]}

\def\oiii{[O\,{\sc iii}]}

\def\H2{H$_2$}
\def\p1{Paper~I}

\def\kms {$\rm km\,s^{-1}$}

\def\MS{\,M$_{\odot}$}

\title[Feeding vs. Feedback in NGC~2110]{Feeding Versus Feedback in AGN from Near-Infrared IFU Observations XI: NGC~2110}

\author[Diniz, Riffel, Storchi-Bergmann \& Winge]{Marlon R. Diniz$^{1}$\thanks{E-mail:
diniz.mr@gmail.com}, Rogemar A. Riffel$^{1}$, Thaisa Storchi-Bergmann$^{2}$ and \newauthor Claudia Winge$^{3}$\\
$^{1}$ Universidade Federal de Santa Maria, Departamento de F\'\i sica, Centro de Ci\^encias Naturais e Exatas, 
97105-900, Santa Maria, RS, Brazil\\
$^{2}$ Universidade Federal do Rio Grande do Sul, Instituto de F\'\i sica, CP 15051, Porto Alegre 91501-970, RS, Brazil\\
$^{3}$ Gemini Observatory, c/o Aura, Inc., Casilla 603, La Serena, Chile
}
\begin{document}

\pagerange{\pageref{firstpage}--\pageref{lastpage}} \pubyear{2013}

\maketitle

\label{firstpage}

\begin{abstract}

We present a two-dimensional mapping of the gas flux distributions, as well as of the gas and stellar kinematics in the inner 220\,pc of the Seyfert galaxy NGC\,2110,
using K-band integral field spectroscopy obtained with the Gemini NIFS at a spatial resolution of $\approx24\,$pc 
and spectral resolution of  $\approx40\,$\kms. 
The \H2\,$\lambda\,2.1218\,\mu$m emission extends over the whole field-of-view and is attributed to heating by X-rays from the AGN and/or by shocks, while the Br$\gamma$ emission is restricted to a bi-polar region extending along the South-East--North-West direction. 
The masses of the warm molecular gas and of the ionized gas are $M_{\rm H_2}\approx1.4\times10^3\,{\rm M_{\odot}}$ and $M_{\rm H\,II}\approx1.8\times10^6\,{\rm M_{\odot}}$, respectively.
The stellar kinematics present velocity dispersions reaching 250\,km\,s$^{-1}$, and a rotation pattern reaching an amplitude of 200\,km\,s$^{-1}$. The gas velocity fields present a similar rotation pattern but also additional components that we attribute to inflows and outflows most clearly observed in the molecular gas emission. The inflows are observed beyond the inner 70\,pc and are associated to a spiral arm seen in blueshift to the North-East  and another in redshift to the South-West. We have estimated a mass inflow rate in warm molecular gas of 
$\approx{\rm 4.6\times10^{-4}\,M_{\odot}\,yr^{-1}}$.
Within the inner 70\,pc, another kinematic component is observed in the \H2 emission that can be interpreted
as due to a bipolar nuclear outflow oriented along the East-West direction, with a mass-outflow rate
of $\approx{\rm 4.3\times10^{-4}\,M_{\odot}\,yr^{-1}}$ in warm \H2.
\\
\hfill{\bf Keywords}: galaxies: individual (NGC\,2110) -- galaxies: Seyfert -- galaxies: ISM -- infrared: galaxies -- galaxies: kinematics and dynamics.

\end{abstract}

\section{Introduction}

A necessary condition for the onset of nuclear activity in galaxies is the presence of gas to feed the nuclear Supermassive Black Hole (hereafter SMBH) \citep{heckman14}. Evidence for the presence of large quantities of gas in the inner few hundred parsecs of nearby active galaxies has been indeed found in a number of imaging studies \citep[e.g.][]{malkan98}. These studies reveal the presence of dusty structures, in the form of compact nuclear disks, filaments and spirals. In \citet{simoes}, it was found that, in the particular case of an early-type sample of active and non-active galaxies, all the active galaxies had dusty structures in the inner few hundred parsecs, while only 25\% of the non-actives had such structures, implying that the dust was a tracer of the gas feeding the active nucleus. The excess of dust in the active galaxies relative to the non-active ones was later confirmed by a study using Spitzer observations to calculate the dust mass of these regions \citep{martini13}. 
As the inferred gas mass around the active galactic nuclei (AGN) are much larger than necessary to feed the SMBH inside,
these structures are reservoirs of material where new star formation may be triggered.

This connection between the presence of dusty spirals and nuclear activity in galaxies suggests that the gas might be streaming along these
 structures toward the central region to feed the SMBH inside \citep{ferrarese05,somerville08,kormendy13}.
The dust mass at kpc scales in AGNs is estimated to be in the range $10^5 - 10^7$\,M$_{\odot}$ \citep{simoes,martini13}
and the associated large amounts of molecular gas ($10^7$ to $10^9$\,M$_{\odot}$) reinforces the importance
of searching for signatures of molecular-gas inflows  within the inner kiloparsec.

On the other hand, recent studies have also found massive outflows of molecular gas \citep{sakamoto10,aalto12,veilleux13}, in particular,
in  Luminous Infrared Galaxies (LIRGS) and Ultra Luminous Infrared Galaxies (ULIRGS).

Observations of nearby active galaxies using spatially resolved IFS provide constraints for the
feeding and feedack processes.
The study of these processes in nearby AGNs have been carried out using IFS at near-IR bands, which contain emission lines from molecular
 (tracer the AGN feeding) and ionized (tracer of the AGN feedback) gas. Most of the observations are performed with adaptive optics using 8-10~m 
telescopes at spatial resolutions ranging from a few parsecs to tens of parsecs. The main results of these studies are:
(i) In general, the molecular and ionized gas present distinct flux distributions and kinematics.
The \H2 emission is dominated by rotational motion and shows lower velocity dispersion values than the ionized gas. 
From the gas excitation and kinematics, we have found that the emitting  \H2  gas (hot molecular gas, as it originates in gas at $\approx$\,2000\,K, \citet{sb09}) is located in the plane of the galaxy, presenting a velocity field dominated by rotation. 
In some cases, in the inner $\approx\,30$\,pc of the galaxy,  compact molecular disks \citep{hicks13,mazzalay14,mrk1066c,mrk766} and gas inflows along nuclear spirals \citep[e.g.][]{n4051,mrk79} are found. 
The inflow rates in cold molecular gas range from a few tenths
to a few solar masses per year \citep{n4051,n7582,davies09,sanchez09,mrk1066c,mrk79}.
(ii) Although the ionized gas kinematics frequently presents a rotating disk component similar to that seen in \H2 for most galaxies, it shows a large contribution from an outflowing component  for some objects
\citep{riffel06,n7582,mrk79,ngc5929,mrk1066c,mrk1157,sb10,mrk766,fausto14}.	
These outflows are most easily seen in the \feii\,1.257\,$\mu$m and 1.644$\mu$m emission lines, present in the near-IR spectra. 
In \citet{fausto14} it is shown that these lines seem to be better tracers of AGN outflows than the previously used [O {\sc iii}]\,$\lambda$\,5007\,$\AA$ emission line.
This is due to the fact that the \feii\ emission is more extended than that
of the [O\,{\sc iii}], reaching out to the partially ionized region (beyond the fully ionized region).
Mass-outflow rates in ionized gas have been estimated to be in the range from $0.05$ to ${\rm 6\,M_{\odot}\,yr^{-1}}$ \citep{mrk1157,sb10,mrk79,fausto14}.

In this work we present adaptive optics assisted K-band IFS of the inner
$3''\times3''$ of the galaxy NGC\,2110 obtained with Gemini NIFS. The results shown here are part of a larger project of the AGNIFS (AGN Integral Field Spectroscopy)
group to map inflows and outflows in a complete sample of AGN, selected by their X-ray emission.
NGC\,2110 is an early type S0 galaxy which hosts a Seyfert 2 nucleus, 
located at a distance of 30.2\,Mpc, assuming a redshift $z=0.007789$ (NED\footnote{NASA/IPAC extragalactic database}), where $1''$ corresponds to 146\,pc at the galaxy. 
Due to its proximity, it has been already observed in several  wavebands, from radio to X-rays. 
NGC\,2110 is one of the brightest nearby Seyfert galaxies in  hard X-rays (2-10\,keV) and is classified as a narrow-line X-ray galaxy \citep{bradt78}.
Its luminosity is comparable to that observed for Seyfert 1 nuclei and soft X-ray emission was observed extending up to $4''$ to the north of the nucleus \citep{weaver95}. Dust lanes and asymmetries are seen up to $4''$ west of the nucleus in optical \citep{malkan98} and near-IR \citep{quillen99} broad band images obtained with the Hubble Space Telescope (HST). \citet{sb99} obtained long-slit near-IR spectroscopy of the inner $\sim$\,10\arcsec of NGC\,2110 along the major and minor axes of NGC\,2110 finding strong emission of both \feii\ and \H2, as well as signatures of distinct kinematics for these two species. Excitation by X-rays was favored for \H2, while shocks were concluded to be also important for excitation of the \feii.

Observations in radio reveal extended emission in a S-shaped structure,
with an extent of $\approx4''$ along the North-South direction
\citep{ulvestad, nagar99}. Circumnuclear extended gas emission was studied by \citet{evans} using Chandra, HST and VLA observations.
Long-slit spectroscopic observations of the inner $10''$ \citep{delgado02, ferruit04} 
have shown that the gas velocity field is asymmetric relative to the nucleus at distances larger than $1''$,
while within the inner $1''$ a blueshift excess is observed, 
suggesting the presence of a nuclear outflow.
\citet{rosario} have shown that the outflow is oriented at a position angle (PA) 
offset by $\approx40^{\circ}$ from the PA of the radio jet and 
suggested that the gas is ionized by the central AGN and not by shocks.

The gas kinematics was first investigate by \citet{wilson85} and \citet{wilson85b}, who reported a rotation pattern in the ionized gas kinematics, similar to those observed for other spiral galaxies, suggesting that the gas is in rotation in the galaxy disk. They found that the rotation center was located $1.7''$ south of the photometric nucleus and observed asymmetric profiles in the \oiii  \,\, to the southeast. In the long-slit near-IR  spectroscopic observations of \citet{sb99}, they also found a displacement of the kinematic nucleus, but smaller than the optical, thus attribuited at least partially  to obscuration, which is smaller in the near-IR than in the optical. Recent optical IFS with the Gemini Multi-Object Spectrographs (GMOS) by  \citet{allan} revealed the presence of four 
distinct kinematic components in the emitting gas: a cold gas disk, a warm gas disk,
a compact nuclear outflow and a high latitude cloud identified as ``the northern cloud". They also observed 
excess blueshifts and redshifts in the cold gas disk kinematics relative to rotational motion that were attributed to inflows and outflows.They have shown that the apparent asymmetry of the gas velocity field is due to the superposition of the contribution of the different components; when this is taken into account, there is no more asymmetry in the velocity fields.

Studies of the nuclear stellar population by \citet{delgado01} using optical continuum images
did not show signatures of a young stellar population (age $\leq$ 1\,Gyr). On the other hand, 
\citet{durre14} used near-IR (Z, J and H bands) IFS of the inner 0\farcs8 obtained with the OSIRIS 
instrument at the Keck Telescope at an angular resolution of 70 mas, and  
reported the detection of four star forming clusters at distances smaller than 0\farcs3 from the galaxy nucleus. They also found that the ionized gas (in particular \feii) surrounding the star clusters is being excited by strong outflows
associated with recent star formation and derived star formation rates of 0.12--0.26\,M$_{\odot}$yr$^{-1}$. 
The new findings by \citet{durre14} and \citet{allan}, relative to  previous studies can be attributed to the better spatial coverage of IFS to study the central region of galaxies when compared to long-slit studies. \citet{durre14}  also calculated the SMBH mass using the He\,I velocity profile considering a simple Keplerian rotation in a
region with radius 56\,pc  and found a mass for the SMBH of $\approx4\times10^8$M$_{\odot}$.  
Using the M-$\sigma$ relationship they find a value of
$\approx3\times10^8$M$_{\odot}$ while with the \citet{graham}
updated relationship for elliptical galaxies they derive a
value of $\approx5\times10^7$M$_{\odot}$; a lower value for the SMBH mass was also obtained using the \feii velocity field, probably 
due to the presence of outflows from the clusters.

This paper is organized as follows. In Sec.~\ref{obs} we describe the observations and
data reduction procedures. The results are presented in Sec.~\ref{results} and discussed
in Sec.~\ref{discussion}. We present our conclusions in Sec.~\ref{conclusions}.


\section{Observations and data reduction}\label{obs}

The near infrared observations of NGC\,2110 were obtained with the integral field spectrograph
NIFS \citep{mcgregor03} on the Gemini North telescope in two distinct nights,
under the programme GN-2010B-Q-25. 
In the first night, five exposures of 600 seconds were obtained  for the galaxy,
3 observations with the same exposure time for the sky and 3 exposures of 18 seconds each for the
telluric star. In the second night, only one exposure of 600 seconds for the galaxy 
and for the sky and 4 observations of 15 seconds for the telluric standard star were obtained.

The data reduction was performed using specific tasks contained in the {\sc nifs} package which
is part of {\sc gemini iraf} package. The reduction process followed standard procedures such as trimming
of images, flat-fielding, sky subtraction, wavelength and spatial distortion calibrations.
The telluric absorptions were removed using the spectrum of the telluric standard star
and the flux calibration was performed by interpolation a black body function to the 
spectrum of the telluric standard star.
We combined the individual datacubes into a single datacube using the 
{\it gemcombine} task of the {\sc gemini iraf} package and
used the {\it sigclip} algorithm to eliminate the remaining cosmic rays and bad pixels, using the peak of
the continuum as reference for astrometry for the distinct cubes.
The final data cube covers the inner 3\arcsec$\times$3\arcsec($\sim$440$\times$440\,pc$^2$) region and
contains $\sim3900$ spectra, each spectrum corresponding to an angular
coverage of 0$\farcs$05$\times$0$\farcs$05, which translates into $\sim$7$\times$7\,pc$^2$
at the galaxy.

The spectral resolution of the data is about $3.4\AA$ at $2.3~\mu$m, measured from the full width at  
half maximum (FWHM) of the arc lamp lines and corresponds to a velocity resolution of $~44$\,km\,s$^{-1}$.
The angular resolution is $0\farcs15$, corresponding to $\approx22$\,pc at the galaxy, and was obtained from the FWHM
of a Gaussian curve fitted to the flux distribution of the standard star.
 A similar FWHM is found by fitting the unresolved nuclear K-band continuum  emission, attributed to the emission of the dusty torus. 

For illustration purposes we rebinned each spaxel to 1/3 of its original size (0\farcs05) to smooth the flux and kinematics maps. We applied the same procedure to the continuum image of the standard star and noticed that the effect of this procedure in the measured FWHM is smaller than 5\% and thus the rebinning of the maps does not degrade the spatial resolution of the data significantly.


\section{Results}\label{results}

\begin{figure*}
 \centering
\includegraphics[scale=1]{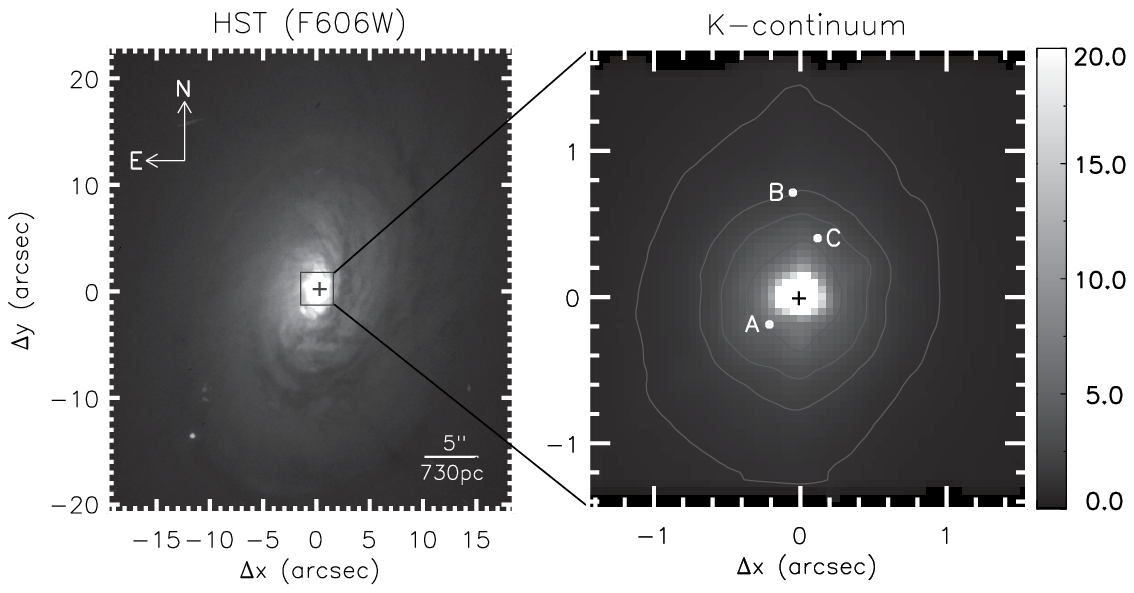}
\includegraphics[scale=1]{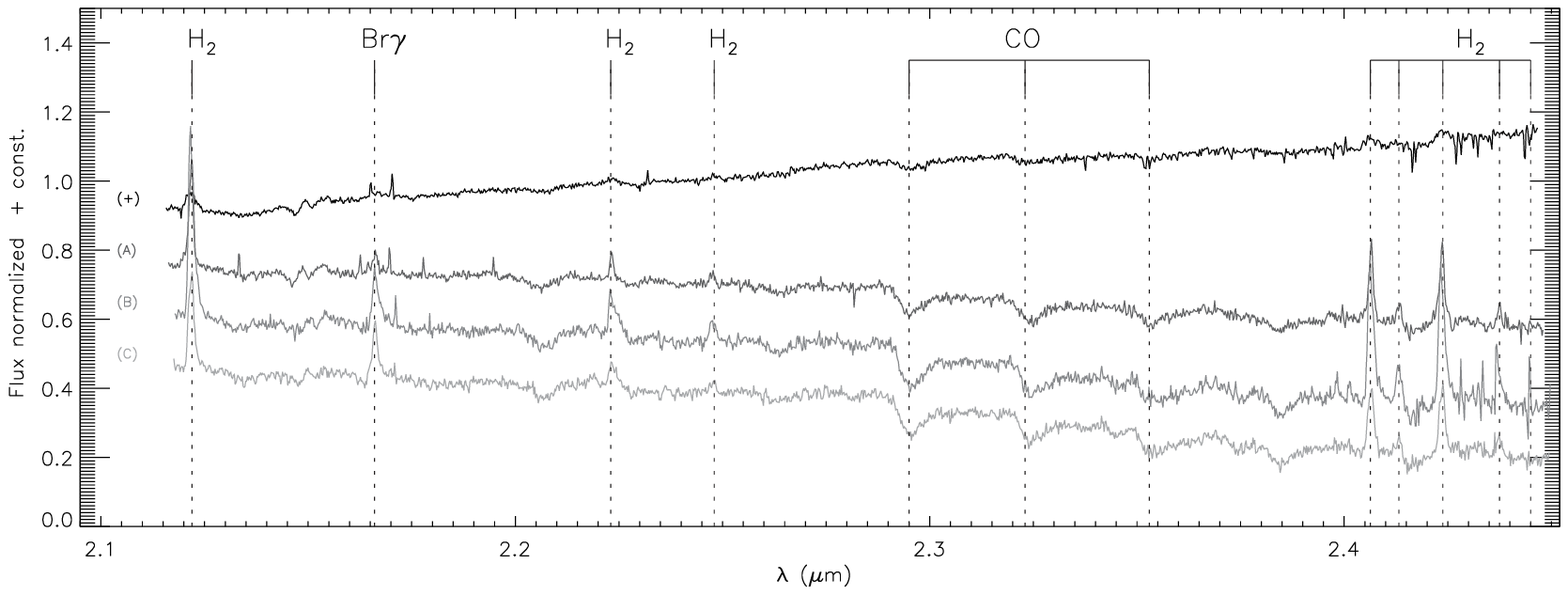} 
\caption{\small The top left panel shows an HST image of NGC\,2110 with the filter F606W,
with the FOV of the NIFS observations indicated by the central square. The top right panel shows
an average continuum image (obtained in the spectral range between $2.25-2.28\mu$m) in flux units of $\rm{10^{-17}\,erg\,s^{-1}\,cm^{-2}}$\,\AA$^{-1}$;
the central cross indicates the position of the nucleus and the other points indicate the positions from which the spectra were extracted.
The bottom panel displays the spetcra in arbitrary flux units per wavelength, which were extracted within a circular aperture with radius of $0\farcs25$ and corrected for Doppler shift.
}
\label{hubble}
\end{figure*}

The left panel of Figure ~\ref{hubble} shows an optical image of NGC\,2110 
obtained with the Hubble Space Telescope (HST) Wide Field Planetary Camera 2 (WFPC2)
through the filter F606W. The 
FOV of the NIFS observations ($3^{\prime\prime}\times3^{\prime\prime}$) is indicated by the gray square.
In the right panel we present a continuum image of the nuclear region obtained from the NIFS
data cube by calculate the average of the fluxes between 2.25 to 2.28~$\mu$m\, a region with no emission lines. 
In the bottom panel we present four spectra, extracted within a circular aperture with radius of $0\farcs25$ at the positions marked in the K continuum image. The position ``+" corresponds to the location of the nucleus, defined as the peak of the continuum emission. 
The position A [($x,y$)=($-$0\farcs2, $-$0\farcs2)] was chosen to represent a typical extranuclear spectrum, 
the position B [with coordinates ($x,y$)=(0\farcs0, 0\farcs7) relative to the nucleus] is located 0\farcs7 north of the nucleus and corresponds to a location where a region og enhanced \H2 line emission is observed (as seen at the top panel of Fig.~3).
Position C [($x,y$)=(0\farcs2, 0\farcs4)] corresponds to a location of enhanced Br$\gamma$ emission at 0\farcs4 northwest of the nucleus.
It can be seen that the extra-nuclear spectra are very similar
(positions A, B and C), while the nuclear spectrum (marked with the '+' sign) is much redder, probably due to emission of the dusty torus.
At most positions the spectra display several emission lines of \H2, Br$\gamma$, as well as the CO absorption-band heads around 2.3~$\mu$m. 
These lines are identified in the spectra and have allowed us to obtain the gas and stellar kinematics, as discussed below.

\subsection{Stellar kinematics}\label{stellar}

We have used the penalised pixel-fitting (pPXF) method of \citet{cappellari04}
to derive the stellar kinematics, by fitting the $^{12}$CO$\lambda\,2.29\mu$m, $^{12}$CO$\lambda\,2.32\mu$m,
$^{13}$CO$\lambda\,2.34\mu$m stellar absorptions
in the K band following the procedure described in \citet{n4051}.
The library of 60 stellar templates of early-type stars \citep{winge09}
was used to obtain the stellar line-of-sight
velocity distribution (LOSVD) at each position.

\begin{figure}
 \centering
\includegraphics[scale=0.9]{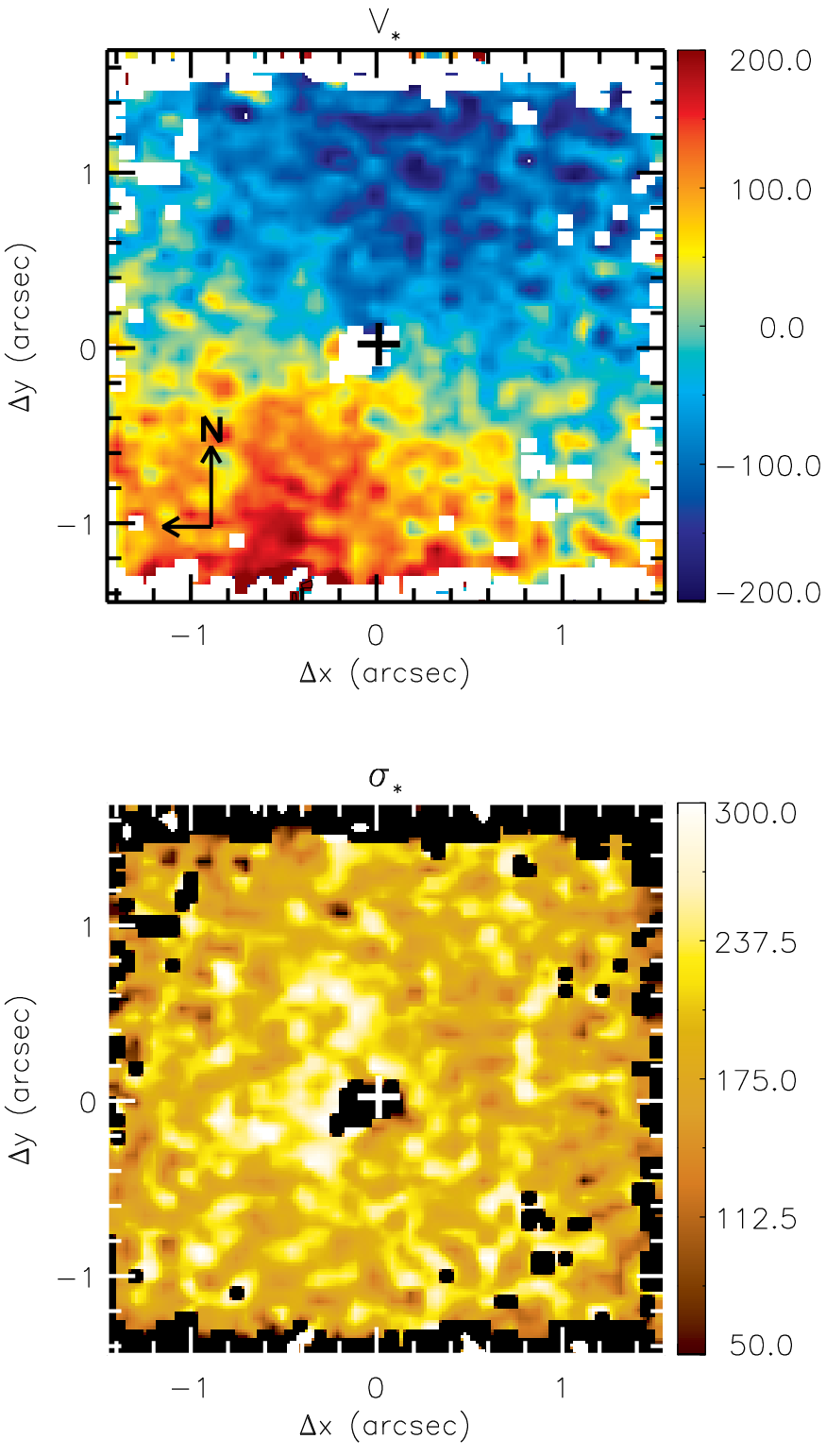}
\caption{\small Stellar kinematic maps obtained from the pPXF fit.
Top: centroid velocity map. Bottom: velocity dispersion map.
The color bar shows the velocity values in units of km\,s$^{-1}$ and the central cross marks the position of the nucleus.}
\label{stel}
\end{figure}

The top panel of Fig.~\ref{stel} presents the velocity (V$_*$) map
from which we subtracted the systemic velocity, obtained by fitting the velocity field as discussed in section~\ref{kinematics-2}.
This map shows a rotation pattern with blueshifts to the North-West and
redshifts to the South-East of the nucleus, with a maximum velocity of $\approx200\,$km\,s$^{-1}$ and
a major axis along the North-South direction at PA\,$\approx\,170^{\circ}$.
The bottom panel displays the stellar velocity dispersion ($\sigma_*$) map 
with values ranging from 50 to 300\,km\,s$^{-1}$.
White regions in the V$_*$ map and black regions in the $\sigma_*$ map correspond
to positions where the signal-to-noise ratio
in the CO bands was not high enough to allow good fits. At the nucleus of the galaxy, the couldn't get good fits due to the 
dilution of the CO absorptions by non-stellar continuum emission. The dilution may be caused by emission of hot dust, as seen in the nuclear spectrum shown in Fig.~\ref{hubble}. 
The uncertainties were derived from the pPXF routine
and are smaller than 30\,km\,s$^{-1}$ for both the V$_*$ and $\sigma_*$ maps.

\subsection{Emission-line flux distributions}\label{flux_distributions}

In order to construct maps for the gas flux distributions, centroid velocities and
velocity dispersions, we fitted the emission lines of \H2 and Br$\gamma$ with Gaussian curves
using a modified version of the emission line profile fitting routine PROFIT \citep{profit}.
This an automated routine written in IDL language and  can be used to fit emission lines with Gaussian curves or Gauss-Hermite Series using the  MPFITFUN routine \citep{mark09} to perform
the nonlinear least-squares fit.  More details of the routine can be found in \citet{profit}. 

\begin{figure}
\begin{center}
    \includegraphics[scale=0.9]{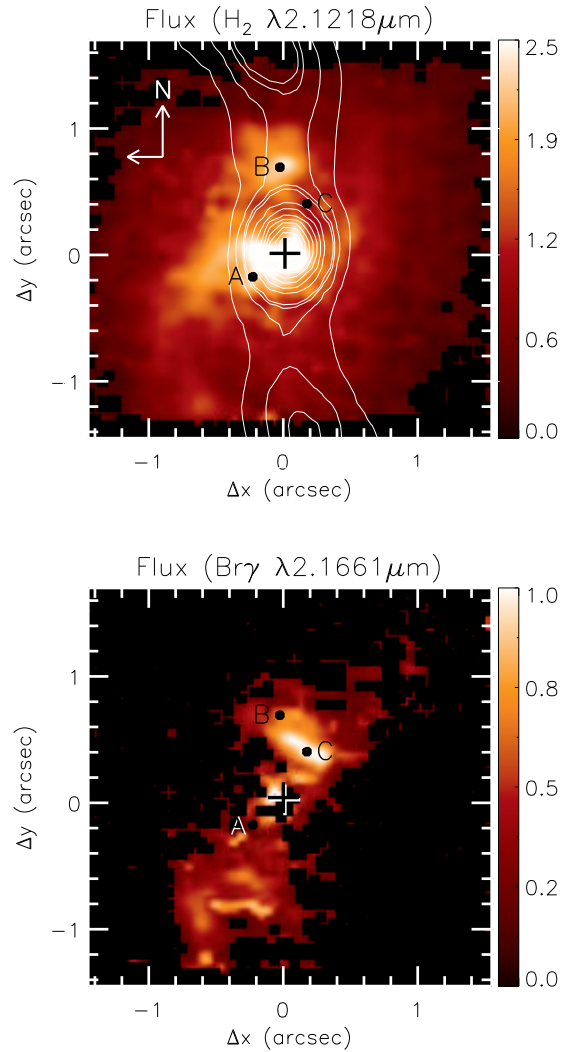}
\caption{\small Top: \H2 flux distribution. Bottom: Br$\gamma$ flux distribution.
The central cross marks the position of the nucleus, the color bars show the range of flux values for each emission line in units of 10$^{-17}$ erg s$^{-1}$ cm$^{-2}$ and 
white contours correspond to the radio emission.}
\label{fluxos}
\end{center}
\end{figure}

The flux distributions for the emission lines H$_2\,\lambda\,2.1218\mu$m and 
Br$\gamma\,\lambda\,2.1661\,\mu$m (in units of $\rm{10^{-17}\,erg\,s^{-1}\,cm^{-2}}$) are shown,
respectively, in the top and bottom panels of Fig.~\ref{fluxos}.
The H$_2\,\lambda\,2.1218\,\mu$m flux map shows emission over the whole FOV, with 
two regions of higher intensity relative to the surrounding regions,
one at the nucleus and other at $\approx0\farcs7$ to the North of it. 
The flux distributions for other 7 emission lines of \H2
(${\rm H_2\lambda\,2.22344\,\mu m}$,
${\rm H_2\lambda\,2.24776\,\mu m}$,
${\rm H_2\lambda\,2.40847\,\mu m}$,
${\rm H_2\lambda\,2.41367\,\mu m}$,
${\rm H_2\lambda\,2.42180\,\mu m}$,
${\rm H_2\lambda\,2.43697\,\mu m}$ and
${\rm H_2\lambda\,2.45485\,\mu m}$)
are similar to that of H$_2\,\lambda\,2.1218\,\mu$m, presenting however a lower signal to noise ratio
and for this reason they are not shown. 

The Br$\gamma$ flux map shows emission only covering part of the FOV being extended along the South-East--North-West direction,
with the peak flux at $\approx0\farcs4$ North-West of the nucleus.
The uncertainties in flux are smaller than 20\% at most locations for both
emission lines and black regions in Figure~\ref{fluxos} correspond to locations
where the uncertainties are larger than 50\% and no reliable fits were obtained.

Tab.~\ref{tab-k} displays the fluxes for the detected emission lines
at positions A (next to the nucleus), B (the northern blob of enhanced emission in the \H2 flux map of Fig.~\ref{fluxos}) and C (corresponding
to the peak of the Br$\gamma$ emission at $0\farcs4$ North-West of the nucleus in the flux map of Fig.~\ref{fluxos}), integrated within
a circular aperture of radius $0\farcs25$. 

\begin{table}
\scriptsize
\centering
\caption{Emission-line fluxes (units of $\rm{10^{-16}\,erg\,s^{-1}\,cm^{-2}}$) for the A, B and C regions identified in Fig.~\ref{fluxos}
integrated within a circular aperture with radius of $0\farcs25$. }
\vspace{0.3cm}
\begin{tabular}{p{1.0cm}ccccc} 
\hline
\\[-0.25cm]
$\lambda_{0}$ ($\mu$m) 			&  ID & Nuc  & 	A ($0.2''$\,SE) 	&  B ($0.7''$\,N)  & C ($0.45''$\,NW) \\
\\[-0.25cm]
\hline
\\[-0.2cm]
2.12183 & 	H$_{2}$ 1-0S(1)		&17.3$\pm4.2$ &  5.55$\pm0.86$ &5.95$\pm0.93$ & 4.14$\pm0.19$ \\ \\[-0.3cm] 
2.16612 &	HIBr$\gamma$ 		& 	-      & 1.12$\pm0.10$ &1.33$\pm0.19$ & 2.46$\pm0.59$  \\ \\[-0.3cm] 
2.22344 &	H$_{2}$ 1-0 S(0)	& 	-      & 1.14$\pm0.11$ &2.14$\pm0.41$ & 1.41$\pm0.18$ \\ \\[-0.3cm]
2.24776 &	H$_{2}$ 2-1 S(1)	& 	-      & 0.83$\pm0.08$ &0.68$\pm0.05$ & 0.48$\pm0.05$ \\ \\[-0.3cm] 
2.40847 &	\H2 1-0\,Q(1)		& 	-      & 3.23$\pm0.48$ &4.52$\pm0.57$ & 3.34$\pm0.26$ \\ \\[-0.3cm]
2.41367 & 	\H2 1-0\,Q(2)		& 	-      & 0.84$\pm0.10$ &1.54$\pm0.57$ & 0.90$\pm0.33$ \\ \\[-0.3cm]
2.42180 & 	\H2 1-0\,Q(3)		&       -      & 4.24$\pm0.36$ &4.45$\pm0.48$ & 3.38$\pm0.20$ \\ \\[-0.3cm]
2.43697 & 	\H2 1-0\,Q(4)		&       -      & 0.79$\pm0.07$ &1.34$\pm0.18$ & 0.67$\pm0.18$ \\ \\[-0.3cm]
2.45485	& 	\H2 1-0\,Q(5)		&       -      & 2.86$\pm0.26$ &5.04$\pm0.50$ & 1.93$\pm0.77$ \\ \\[-0.3cm] 
\hline 
\end{tabular}
\label{tab-k}
\end{table}


\subsection{Centroid velocity and velocity dispersion maps for the emitting gas}\label{centroid}

\begin{figure}
\begin{center}
    \includegraphics[scale=0.9]{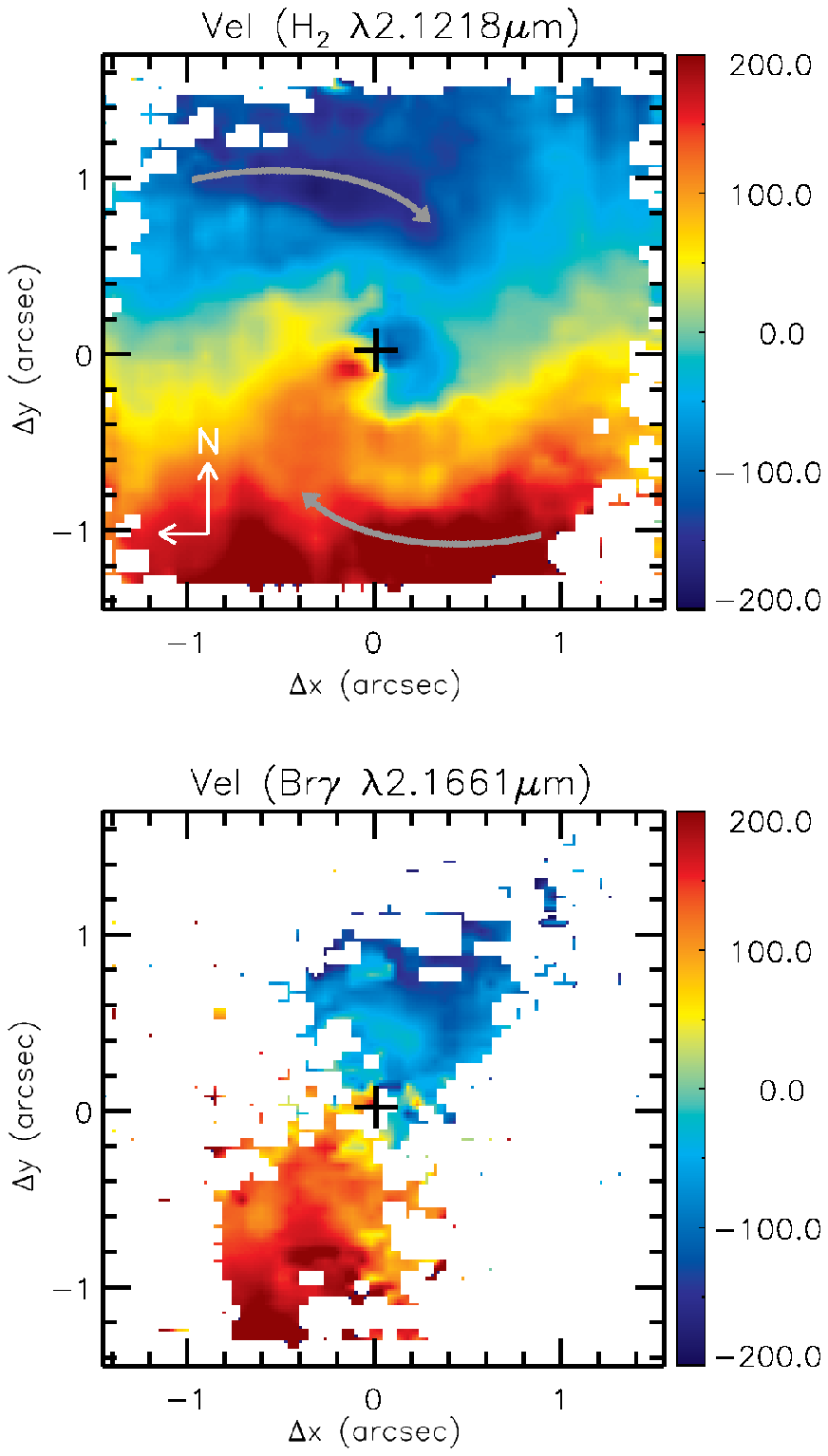}
\caption{\small Top panel: velocity field of the \H2 emitting gas. Bottom panel: velocity field of the
Br$\gamma$ emitting gas, restricted around PA$\approx -30^{\circ}/150^{\circ}$. 
The color bars are in units of km\,s$^{-1}$ and the mean uncertainty in velocity is less than 10\,km\,s$^{-1}$. 
The spirals delineated by the gray curves shown in the \H2 velocity map mark the location of the inferred inflows discussed in Sec.~\ref{feeding}.
In both maps, regions with errors higher than 50\% in flux were masked out,
as for the flux map.}
\label{vels}
\end{center}
\end{figure}

The top panel of Fig.~\ref{vels} shows the centroid velocity map 
for the molecular gas in units of km\,s$^{-1}$. The systemic velocity, as determined from the fit of the stellar velocity field, 
was subtracted and the cross indicates the position of the nucleus.
The \H2\ velocity field shows a rotation pattern
with blueshifts to the North (thus approaching)
and redshifts to the South (thus receding), with a velocity amplitude of
$\approx\,200\,$km\,s$^{-1}$. But, although presenting similar rotation to that observed in the
stellar velocity field (Fig.~\ref{stel}), the \H2 velocity field presents additional components.
Two structures resembling spiral arms are observed: one to the North in blueshift
and the other to the South in redshift. Closer in, excess blueshifts are observed to the West and excess redshifts to the
East, at distances from the nucleus smaller than $0\farcs5$ (73\,pc). The dark gray curves with an arrow represent the gas flowing to the central region along nuclear spirals.

\begin{figure}
\begin{center}
    \includegraphics[scale=0.9]{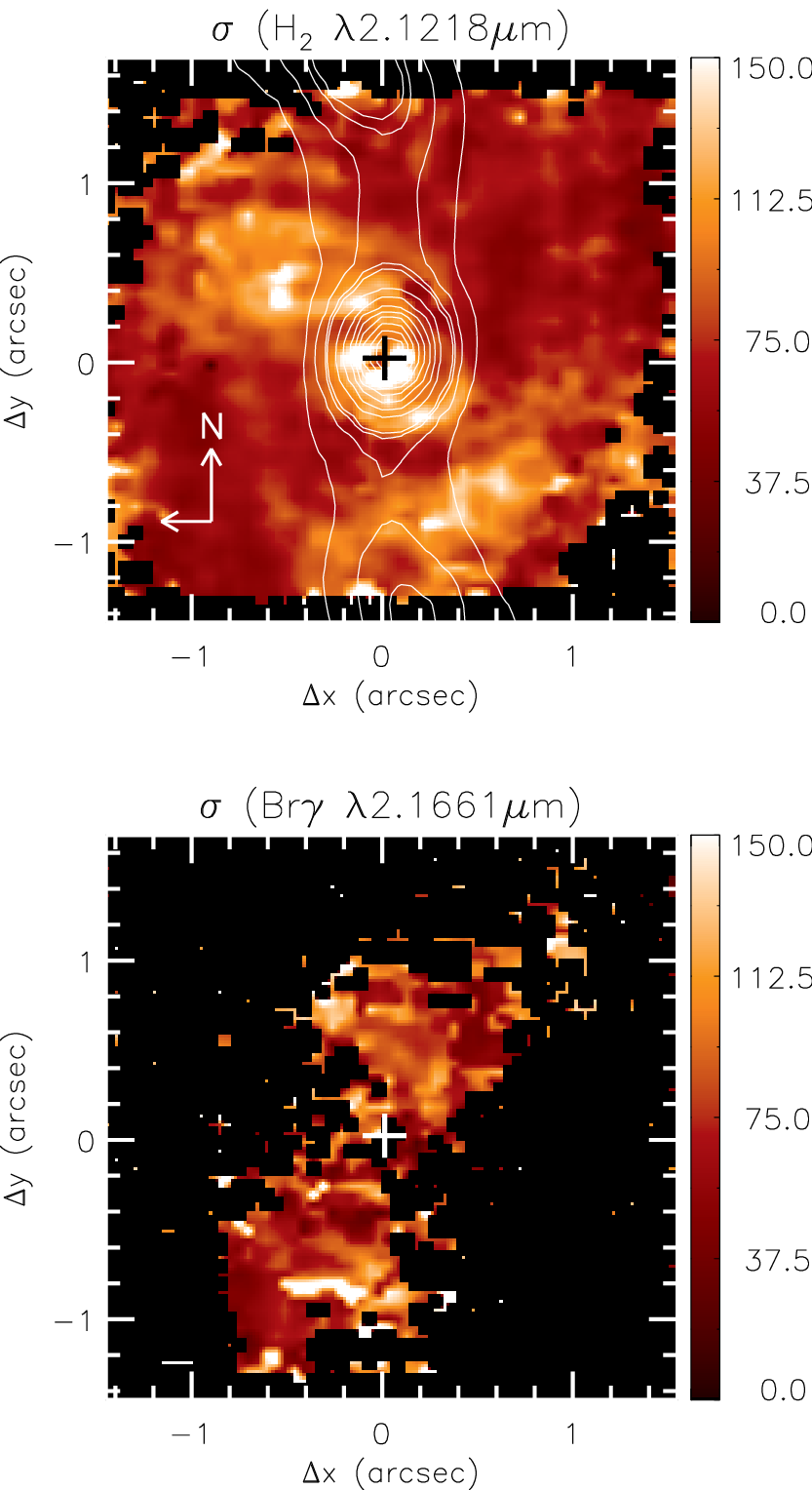}
\caption{\small Top: velocity dispersion map for \H2 with 
contours of the 3.6-cm Very Large Array (VLA) radio map superimposed.
Bottom: velocity dispersion map for Br$\gamma$. Both maps are in units of km\,s$^{-1}$.}
\label{disp}
\end{center}
\end{figure}

Figure~\ref{disp} shows the velocity dispersion ($\sigma$) maps for the \H2
and Br$\gamma$ emitting gas. Both maps have been corrected for the instrumental broadening. 
The top panel shows the $\sigma$ map for the
H$_2\lambda\,2.1218\mu\,$m emission line, with uncertainties of $\approx\,10$km\,s$^{-1}$.
High velocity dispersions of $\approx\,150\,$km\,s$^{-1}$ are seen
in a region extending along PA$\approx\,-45^{\circ}$ from North-East to South-West
and lower values  of $\approx\,60$\,km\,s$^{-1}$ are observed in all other regions.

The Br$\gamma$ $\sigma$ map is shown in the bottom panel of Fig.~\ref{disp},
with mean uncertainties of 15\,km\,s$^{-1}$ and presents somewhat higher
$\sigma$ values than those observed for the \H2 in the same regions.
At most locations $\sigma\approx\,100$\,km\,s$^{-1}$ are observed in the regions
where the \H2 $\sigma$ is $\approx\,60$\,km\,s$^{-1}$.

\subsection{Channel Maps}

\begin{figure*}
    \includegraphics[scale=0.6]{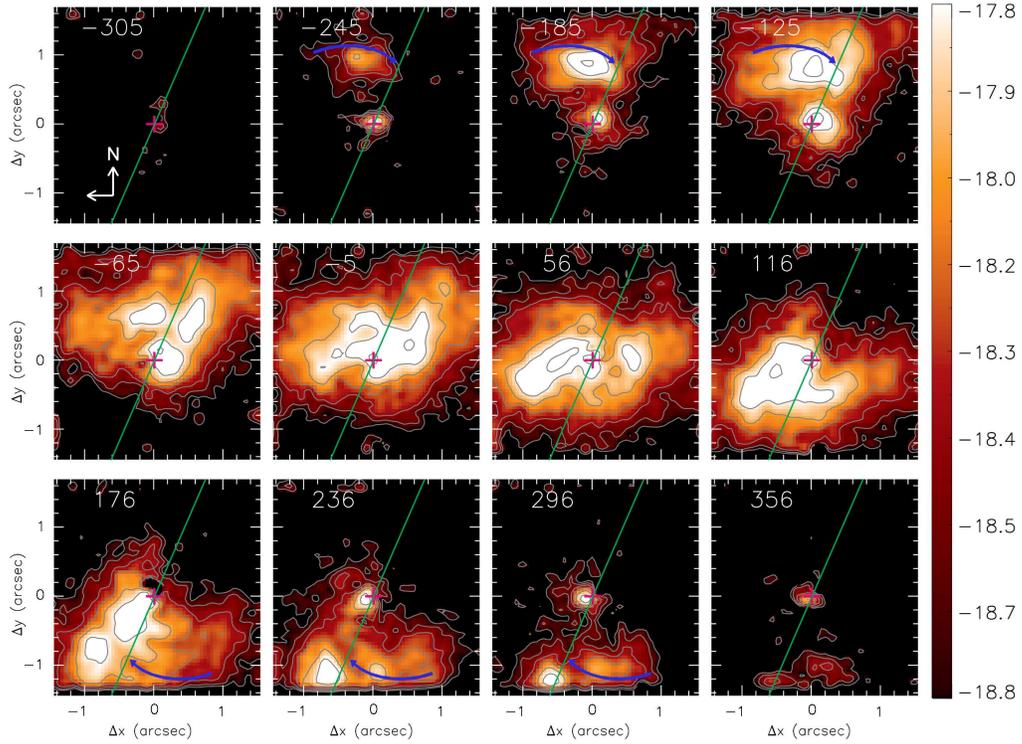}
\caption{\small Velocity channels along the \H2 emission line in $\approx\,60\,$km\,s$^{-1}$  bins (larger than velocity resolution of the data FWHM $\approx\,44$\,km\,s$^{-1}$) centered on the
velocities indicated in each panel. The central cross marks the position of the nucleus (peak of continuum emission), the
intensities are represented in the color scale to the right in logarithmic units and the green line corresponds to the line of nodes. 
The spiral arms are delineated by the blue curves shown in the \H2 channel maps indicating the location of the inferred inflows discussed in Sec.~\ref{feeding}.} 
\label{channel-h2}
\end{figure*}

\begin{figure*}
    \includegraphics[scale=0.6]{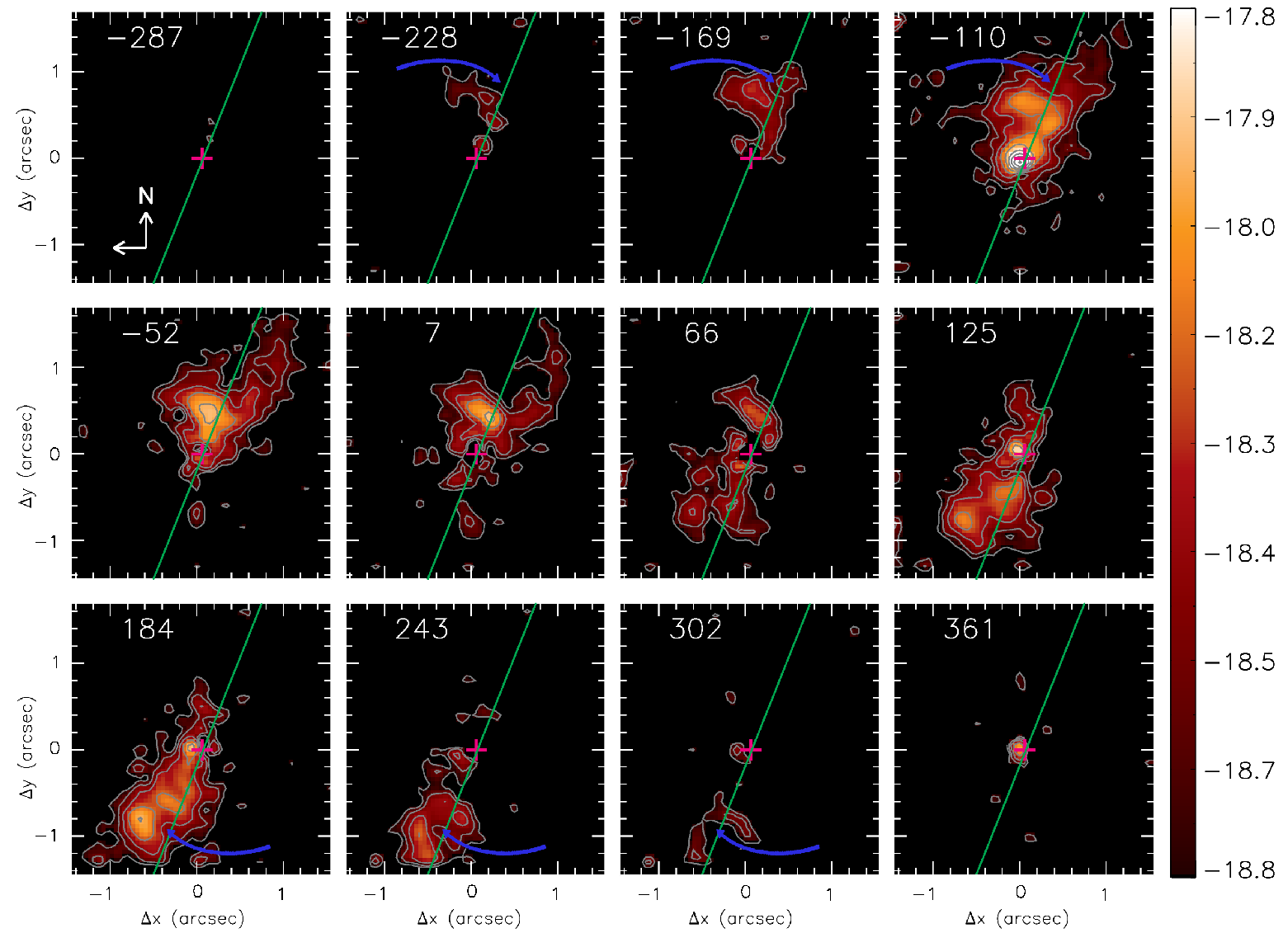}
\caption{\small Same as Fig.~\ref{channel-h2} for Br$\gamma$.}
\label{channel-brg}
\end{figure*}

Figures ~\ref{channel-h2} and ~\ref{channel-brg} show channel maps along the H$_2\,\lambda\,2.1218\mu$m and Br$\gamma$
emission line, respectively. In each panel,  the flux distributions have been
integrated within velocity bins of 60\,km\,s$^{-1}$, corresponding to two spectral pixels.
The flux distributions are shown in logarithmic units and the central velocity of each panel (relative to the systemic velocity of the galaxy)
is shown in the top-left corner. The central cross marks the position of the nucleus.

Fig.~\ref{channel-h2} shows that the molecular hydrogen 
presents emission from gas with velocities in the range from $-300$ to $+350$ km\,s$^{-1}$.
The highest blueshifts are seen at $~1''$ North of the  nucleus, and at a small region surrounding
the nucleus, while the highest redshifts are observed predominantly to the South of the nucleus
at distances $~1''$ from it and between the nucleus and $0\farcs2$ to the East.
At lower velocities (from $\approx\,-100$ to $+100$\,km\,s$^{-1}$),
the \H2 emission spreads over a larger part of the FOV with the highest intensity
structure moving from North-West to South-East as the central velocities change from blueshifts 
to redshifts.

The channel maps across the Br$\gamma$ emission line profile are shown in Fig.~\ref{channel-brg}.
The highest blueshifts and redshifts are observed to the North-West and South-East of the nucleus,
respectively. There is a trend similar to that observed for \H2 for velocities between $\approx\,-100$ and +\,170\,km\,s$^{-1}$: as the velocities change from blueshifts to redshifts,
the highest intensities move from North-West to South-East of the nucleus.

\section{Discussion}\label{discussion}

\subsection{Stellar kinematics}\label{kinematics-2}

\citet{delgado02} presented a study of the stellar kinematics of the central region
of NGC\,2110 based on optical long-slit observations along PA=6$^{\circ}$
of the Ca {\sc ii} triplet ($\lambda\lambda 8498, 8542, 8662$) absorption lines.
They measured a velocity amplitude of $\approx\,180\,$km\,s$^{-1}$, with the
the line of nodes oriented along PA$=163^{\circ}$ and velocity dispersion of 
 $\approx\,260\,\pm\,20$\,km\,s$^{-1}$.  This $\sigma$ value agrees, within the uncertainties, with that obtained
by \citet{nelson} ($\sigma\,=\,220\,\pm\,25$\,km\,s$^{-1}$) using Mg\,$b$ lines.

\citet{ferruit04} obtained the stellar kinematics from Mg and Fe absorptions lines in the optical 
with the integral field spectrograph OASIS of the Canada--France--Hawaii Telescope (CFHT)
in the spectral range $4760-5560\AA$ over a $4''\times4''$ FOV. They found a velocity amplitude of 
$\approx\,200$\,km\,s$^{-1}$ with the orientation of the line of nodes being consistent with the 
orientation of the photometric major axis of the galaxy -- PA$=163^{\circ}$.
The authors also fitted the near-IR CO absorption band heads from long slit spectroscopy
obtained with the NIRSPEC instrument of the Keck-II Telescope. 
They found a good correspondence between the NIRSPEC and OASIS measurements.

In order to compare our stellar velocity field with those obtained in the previous studies described above, 
we have fitted a rotating disk model to the stellar velocity field of Fig.~\ref{stel}), assuming 
circular orbits in the plane of the galaxy, as in  \citep{bertola91}:
\begin{equation}
 V(R,\Psi)=V_{s}+\frac{ARcos(\Psi - \Psi_0)sin(i) cos^p\theta}{{R^2[sin^2(\Psi - \Psi_0)+cos^2(i)cos^2(\Psi - \Psi_0)]+c^2_0cos^2(i)}^{p/2}}
\label{model-bertola}
\end{equation}
where $R$ and $\Psi$ are the coordinates of each pixel in the plane of the sky, $V_s$ is the systemic velocity, $A$ is the amplitude of the rotation curve at the plane of the galaxy, $\Psi_0$ is the position angle of the line of nodes, $c_0$ is a concentration parameter, $i$ is the disk inclination relative to the plane of the sky, $p$ is a model fitting parameter. For $p=3/2$ the system has a finite mass, as for a Plummer potential \citep{plum} and for $p=1$ the rotation curve is asymptotically flat.
 
\begin{figure}
\begin{center}
    \includegraphics[scale=0.8]{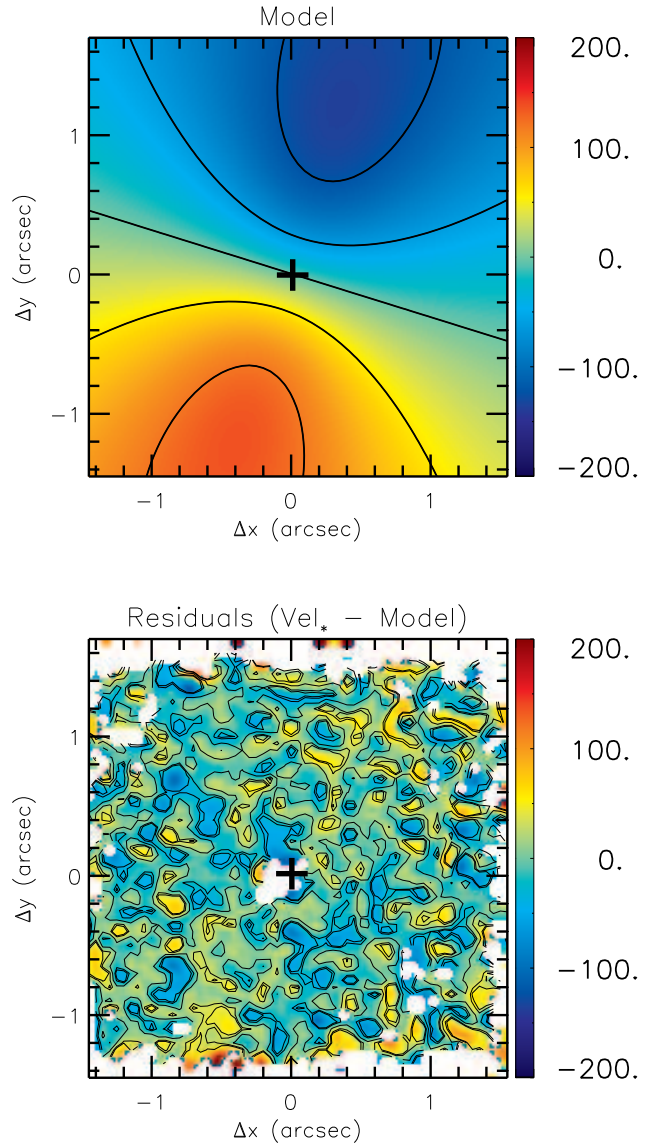}
\caption{\small Rotating disc model for the stellar velocity field (top) and residual
map between the observed and modeled velocities. The cross marks 
the position of the nucleus and the color bar shows the range of velocities in
km\,s$^{-1}$.}
\label{model}
\end{center}
\end{figure}

The top-panel of Fig.~\ref{model} shows the rotating disk model that provided
the best fit of the observed velocities. In the bottom panel we show the residual map, 
obtained from the subtraction of the model from the observed velocities.
Although at some locations the residuals are larger than the calculated uncertainties, the residual map does not show any systematic structure, and presents residuals larger than 50\,\kms only for 13\% of the spaxels. We thus conclude that the model is a good representation of the stellar velocity field.
The parameters derived from the fit are: 
systemic velocity corrected to the heliocentric reference frame
$V_s=2332\,\pm\,15\,$km\,$s^{-1}$, 
$\Psi_0=162^{\circ}\,\pm\,0.6^{\circ}$,
$A=306\,\pm\,10.5\,$km\,$s^{-1}$,
$i=43^{\circ}\,\pm\,2.5^{\circ}\,$, 
$c_0=0.9\,\pm\,0.04$\,arcsec. 
The location of the kinematical center was fixed at
$X_0=Y_0=0$ arcsec and $p$ was limited to range between 1 and 1.5, which is the range of values expected for galaxies \citep{bertola91}. The best fit was obtained with $p=1.5$. 
The systemic velocity and the inclination
of the disk are in good agreement with those listed  in the NED database.
The orientation of the line of nodes agrees with the one quoted in \citet{delgado02}.

We can also compare our results with those obtained for the gas kinematics by \citet{allan} using the GMOS IFU for the inner 1.1$\times$1.6\,kpc$^2$ of NGC\,2110. 
They identified four components in the emitting gas named as: (1) warm gas disk
($\sigma=100-220$\,\kms); (2) cold gas disk ($\sigma=60-90$\,\kms); (3) nuclear component ($\sigma=220-600$\,\kms) and (4) northern cloud ($\sigma=60-80$\,\kms). They fitted the velocity field of the disk components by a rotation model similar to the one used here. For the cold component they found  
$V_s=2309\,\pm\,10\,$km\,$s^{-1}$,  
$\Psi_0=171^{\circ}\,\pm\,1^{\circ}$ and 
$i=39^{\circ}\,\pm\,1^{\circ}\,$. For the warm component they found $V_s=2308\,\pm\,11\,$km\,$s^{-1}$,  
$\Psi_0=169^{\circ}\,\pm\,1^{\circ}$, keeping the inclination fixed to the same value of the cold disk. The parameters for both disk components are similar to those we obtained for the stellar kinematics, as well as the amplitude of the velocity field in the inner 1\farcs5 obtained for the optical emitting gas.

The average stellar velocity dispersion obtained from the measurements shown in Fig.~\ref{stel} 
is $\sigma_ {\star}= 200\,\pm\,20\,$km\,s$^{-1}$, which is somewhat smaller from those obtained by \citet{delgado02} and \citet{nelson} using optical absorption  lines as mentioned above.   
However, \citet{rif2014} showed that the $\sigma_*$ obtained using CO bandheads
is systematically smaller than the values obtained from the fitting of optical absorption
lines with a mean fractional difference ($\sigma_{frac}=(\sigma_{CO}-\sigma_{opt})/\sigma_{opt}$) of 14.3\%, with some galaxies showing much higher
discrepancies (as NGC~4569, with a fractional difference of 40\%).
Integrating the K-band spectra over the whole field of view and using the pPXF code, as done in Sec.~\ref{stellar} we obtain $\sigma_*=230\pm\,4$\,\kms. 
Using the value of \citet{delgado02} we find a fractional difference  of $-$11\%, while using the value of \citet{nelson} we get a difference of 4\%.
These differences are  of the order of the discrepancies found in \citet{rif2014}.

Using this result in the $M_{\bullet} - \sigma_{\star}$ relation \citep{graham} we obtain
a mass for the SMBH of
$M_{\bullet} = 2.7\,^{+3.5}_{-2.1}\,\times\,10^{8}\,$M$_{\odot}$,
which agrees with value found by \citet{Moran07} 
using a value for the velocity dispersion measured in optical observations. 
The  $M_{\bullet}$ we have found in this work is also consistent with that of  
\citet{durre14} obtained using the kinematics of the ionized gas (He\,{\sc i}) to derive the mass enclosed in the inner 56\,pc ($\approx\,4\times10^8{\, \rm M_\odot}$).

\subsection{Gas excitation}\label{disc-excitation}

The \H2 molecule can be excited by processes such as heating of the gas by shocks
\citep{hollenbach89} or by X-rays \citep{maloney96}, known as thermal processes,  and by  non-thermal processes such as absorption of soft-UV \citep{black87}
and far-UV at clouds with density larger than $10^4\,$cm$^{-3}$ \citep{sternberg89}. 
The origin of the \H2 emission in AGN has been discussed in several recent studies,
based on long-slit \citep{sb99, reunanen,davies05} and integral field spectroscopy \citep{riffel06,sanchez09,sb09,mazzalay10}.
In the K band there are several emission lines from \H2, that can be used to study their origin.
We used the ratio \H2$\lambda$2.2477/$\lambda$2.1218 to assess if the \H2 emission is due to a thermal or non-thermal process
at A and B locations in Fig.~\ref{fluxos}. 
For thermal processes, typical values are $\approx\,0.1-0.2$, while for non-thermal processes,
typical values are $\approx\,0.55$  (e.g. \citet{mouri94,reunanen,sb09}).
For NGC\,2110 we find \H2$\lambda$2.2477/$\lambda$2.1218$\,\approx0.15$ for the position A and \H2$\lambda$2.2477/$\lambda$2.1218$\,\approx0.11$ for positions B and C,
indicating that the \H2\ is thermalized by shocks and/or by X-ray emission. However the signal-to-noise ratio for the \H2$\lambda$2.2477 emission line was not high enough to further investigate the \H2$\lambda$2.2477/$\lambda$2.1218 ratio at other locations.

The lower vibrational energy states of the \H2 molecule can also be thermalized in high gas densities even if excited by far-UV radiation \citep{sternberg89}, and long-slit spectroscopy of the circumnuclear region of a few active galaxies suggests that the this might be the case of some objects \citep{davies05}. 

Assuming that the \H2 is thermalized, we can find the excitation temperature by fitting the observed fluxes by the following equation \citep{scoville1982}:

\begin{equation}
{\rm log}\left(\frac{F_{\rm i}\lambda_{\rm i}}{A_{\rm i}g_{\rm i}}\right)=
{\rm constant}-\frac{T_{\rm i}}{T_{\rm exc}},
\label{eqtempex}
\end{equation}
where $F_{\rm i}$ is the flux of the $i\,^{th}$ \H2 line,
$\lambda_i$ is its wavelength, $A_i$ is the spontaneous emission coefficient,
 $g_i$ is the statistical weight of the upper level of the transition,
$T_i$ is the energy of the level expressed as a temperature and 
$T_{exc}$ is the excitation temperature. 
This equation assumes a ratio of transitions {\it ortho:para} of 3:1 and
is  valid  only for thermal equilibrium, what seems to be the case for most AGN,
as found in the studies discussed above: non-thermal processes contribute very little
to the \H2 emission, at least in the cases that do not show intense star formation. 
In the case of NGC\,2110, the \H2$\lambda$2.2477/$\lambda$2.1218 ratio discussed above supports that the \H2\ is thermalized.
 
\begin{figure}
\begin{center}
    \includegraphics[scale=0.4]{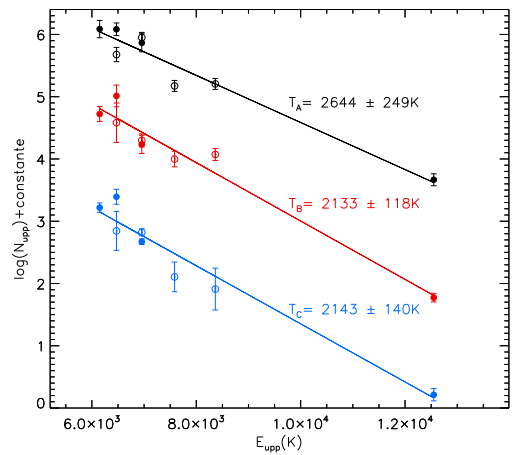}
\caption{\small Relation between 
$N_{\rm upp} = F_{\rm i}\,\lambda_{\rm i}/A_{\rm i}\,g_{\rm i}$
and $E_{\rm upp} = T_i$ for the \H2 emission lines for thermal excitation
at locations A, B and C (from top to bottom) in Fig.~\ref{fluxos}. {\it Ortho} transitions are shown as
filled circles and {\it para} transitions as open circles.}
\label{texc}
\end{center}
\end{figure}

In Fig.~\ref{texc} we plot the
$N_{\rm upp} = F_{\rm i}\,\lambda_{\rm i}/A_{\rm i}\,g_{\rm i}$
(plus an arbitrary constant) $versus$ $E_{\rm upp}=T_i$ 
for the positions A, B and C shown in Fig.~\ref{fluxos},
in which ortho transitions are represented with open symbols and
the best linear fit of equation~\ref{eqtempex}
is shown as a continuous line. This result confirms the thermal excitation. 
Although the uncertainty in the fit might be high, since there is
only one point at the high energy region, it suggests that the vibrational and rotational states of the \H2 molecule are in thermal equilibrium and this favors thermal processes as the main excitation mechanism of the \H2 emission. The derived excitation temperatures for positions A, B and C are $T_{\rm A}=2644\,\pm\,249\,$K, $T_{\rm B}=2133\,\pm\,118\,$K and $T_{\rm C}=2143\,\pm\,140\,$K, respectively.

In order to further investigate the origin of the \H2 emission, we estimate the vibrational temperature ($T_{vib}$) using the \H2\,$\lambda$\,2.1218/2.2477 line ratio in $T_{vib}$=5600/ln(1.355$\times$F\H2$\lambda$2.1218/F\,\H2$\lambda$2.2477) \citep{reunanen} using the fluxes from Table \ref{tab-k}. We find that  $T_{vib}$ is in the range  2260--2540\,K, which are very similar to the values found above for the excitation temperature. The fact that the excitation temperature, obtained using both rotational and vibrational transitions, and the vibrational temperature are similar, supports the assumption above that both rotational and vibrational states are thermalized.

We also investigated the excitation mechanism of the \H2
using the H$_2\,\lambda2.1218\,/Br\gamma$ line ratio.
In starbursts, where the main excitation mechanism is UV radiation,
H$_2\,\lambda2.1218\,/Br\gamma < 0.6$, 
while for AGN, where the \H2 is excited by shocks,
by X-rays from the nucleus or by UV radiation, $0.6 <$ H$_2\,\lambda2.1218\,/Br\gamma < 2$,
as found from long-slit spectroscopy for the nuclear aperture 
\citep{ardila04,ardila05}. Recent studies using long-slit spectroscopy and IFU spectroscopy indicate a broader range of \H2/Br$\gamma$ ratio values for AGN, with values in the range 0.6-6 \citep{rogerio} and 0.5-8 \citep{colina2014}.
Figure~\ref{razao} shows the \H2/Br$\gamma$ line ratio map for NGC\,2110, where the values are between 0.5 and 8 at most locations indicating excitation by the AGN. Approximately in 80\% of the points the values are larger than 2, but 91\% of the points show values between 0.5 and 8, thus still typical of Seyfert galaxies.

Our Br$\gamma$ flux map (Fig.~\ref{fluxos}) presents a fairly similar structure to that seen in the [N{\sc ii}] flux map from \citet{allan} and to that found
for the [O{\sc iii}] emission by \citet{ferruit04}, where the ionized
gas emission is stronger to the North of the nucleus than to the South, although the Br$\gamma$ emission is less extended than both [N{\sc ii}] and  [O{\sc iii}].
The Br$\gamma$ emission is collimated along the direction South-East -- North-West (PA $\approx\,-30^{\circ}/150^{\circ}$) that is the same orientation of the strongest gas emission in the optical as shown in \citet{allan}. The Br$\gamma$ emission can thus be attributed to ionization by the collimated AGN radiation, that illuminates the galaxy disk in that direction. While in \citet{allan} they have detected (much fainter) gas emission beyond this region of strongest emission, we are not able to detect this emission in Br$\gamma$, as this line is much less intense  than the optical lines.

\begin{figure}
\begin{center}
    \includegraphics[scale=0.9]{razao.ps}
\caption{\small Line ratio map H$_2\lambda2.1218$/Br$\gamma$}.
\label{razao}
\end{center}
\end{figure}


\subsection{Mass of the molecular and ionized gas}\label{disc-mass}

The mass of hot \H2 in the inner $3''\times3''$ can be estimated as follows
\citep{scoville1982}:
\begin{eqnarray}
\hspace{0.5cm}  M_{\rm{H_{2}}}=\frac{2\,m_{\rm p}\,F_{{\rm H_{2}}\lambda2.1218}\,4\pi\,d^{2}}{\mathnormal{f}_{v=1,{\rm J=3}}\,A_{S(1)}h\,\nu} \nonumber 
\label{massah1a}
\end{eqnarray}
\begin{eqnarray}
\hspace{1.1cm} \approx\,5.0776 \times 10^{13} \left(\frac{F_{\rm {H_{2}}\lambda2.1218}}{\rm erg\,s^{-1}cm^{-2}}\right) \left(\frac{\rm d}{\rm Mpc}\right)^2 [{\rm M_\odot}],
\label{massah2a}
\end{eqnarray}
where $m_{\rm p}$ is the proton mass,
$F{\rm_{H_{2}}\,\lambda2.1218}$ 
is the flux of H$_2\,\lambda\,2.1218\,\mu$m emission line,
$d$ is the galaxy distance, $h$ is the Planck's constant
and $\nu$ is the frequency of the \H2 line.
We have assumed a typical vibrational temperature of $T\rm{_{vib}=2000}$\,K \citep{sb09},
which implies a population fraction $\rm{\mathnormal{f}_{v=1,J=3}=1.22\times10^{-2}}$
and a transition probability $A_{S(1)}=3.47\times10^{-7}$ s$^{-1}$
\citep{turner1977,n4051}.
Integrating over the whole field of NIFS, the flux is 
$F_{\rm H_{2}\,\lambda\,2.1218}\,\approx\,3.1\,\times\,10^{-14}\,{\rm erg\,s^{-1}\,cm^{-2}}$,
resulting in a hot molecular hydrogen mass of M$_{\rm H_2} \approx\,1.4\,\times\,10^{3}\,{\rm M_\odot}$.
This value is in good agreement with those found for other AGN \citep{sb09,mrk1066a,mrk79}.
The amount of cold molecular gas may be much higher than the value derived above. \citet{dale05} found a ratio between the mass cold ($M_{\rm cold}$)
and hot molecular in the range 10$^{^5}$-10${^7}$ for a sample of six nearby galaxies, while \citet{sanchez09} found a ratio $1\times10^{6}$ for NGC\,1068 and \citet{mazzalay12}
obtained $M_{\rm cold}$/$M_{\rm H_2}$ ranging from 10${^5}$ to 10${^8}$. 
We use here the conversion factor obtained by the most recent study of \citet{mazzalay12} because it includes more galaxies than the previous studies, that can be obtained from the expression:

\begin{equation}
\frac{M_{\rm cold}}{\rm M_{\odot}}\,\approx\,1174\times\left(\frac{L_{\rm {H_{2}}\,\lambda\,2.1218}}{L_{\odot}}\right),
\label{massafria}
\end{equation}
where $L_{\rm H_2}\,\lambda\,2.1218$ is the luminosity of the \H2 line.
We  find $M_{\rm cold} \approx\,9.92\,\times\,10^{8}\,{\rm M_{\odot}}$,
that is about 6 orders of magnitude larger than the mass of hot molecular gas.

The mass of ionized gas can be estimated from the measured flux of
the Br$\gamma$ emission line using the following expression \citep[derived from the equations in][]{osterbrock06}:
\begin{eqnarray}
M_{\rm{H\,II}} \approx\,3 \times 10^{17} \left(\frac{F_{\rm Br\gamma}}{\rm erg\,s^{-1}cm^{-2}}\right) \left(\frac{\rm d}{\rm Mpc}\right)^2 [{\rm M_\odot}],
\label{massabrgamma}
\end{eqnarray}
where $F_{\rm Br\gamma}$ is the integrated flux for the Br$\gamma$ emission line and $d$
(30.2\,Mpc) is the
distance to the galaxy. We assume an electron temperature $T=10^4$\,K and electron density
$N_{\rm e} = 500$\,cm$^{-3}$.
Integrating over the whole IFU field we obtain 
$F_{\rm Br\gamma}=6.5\,\times\,10^{-15}\,{\rm erg\,cm^{-2}\,s^{-1}}$,
resulting in  $M_{\rm H\,II}\,\approx\,1.77\,\times\,10^6\,{\rm M_\odot}$.

\citet{durre14}, through adaptive optics near-infrared integral field spectroscopy with the Keck OSIRIS instrument
found a total mass of ionized gas of $7.5\times10^5~{\rm M_\odot}$ within a FOV of $\approx 2.3 \times 1.8$ arcsec. 
This value is about 40\% of the value we have obtained, but this is consistent with the fact that the area covered by our
FOV is $2.2$ times larger.

The mass of ionized gas derived for NGC\,2110 is thus 3 orders of magnitude smaller than the estimated total mass of molecular
gas, a result that is in agreement with those of previous studies
\citep{ferruit04,sb09,sb10,n4051,mrk79}.

\subsection{Kinematics of the emitting gas}\label{disc-gas}

The velocity fields of the \H2 and Br$\gamma$ emitting gases (Fig.~\ref{vels})
are similar to the stellar one (Fig.~\ref{stel}),
with blueshifts to the North-West and redshifts to the South-East of the nucleus.
However, the \H2 kinematics shows deviations from rotation in regions next to the nucleus
as well as in regions
$\approx\,$1\farcs0 to the North and South, that show a spiral pattern.

Although the inclination of the disk and orientation of the line of nodes obtained from the fit of the stellar velocity field are not that different from those obtained by \citet{allan} for the gas in the optical, there are significant differences between the \H2 velocity field and the stellar velocity field. Several works indeed show that the gas and stellar kinematics are not always consistent with each other in the central region of galaxies \citep{davies14,westoby12,ngc5929a,oasis}.

\begin{figure}
\begin{center}
    \includegraphics[scale=0.9]{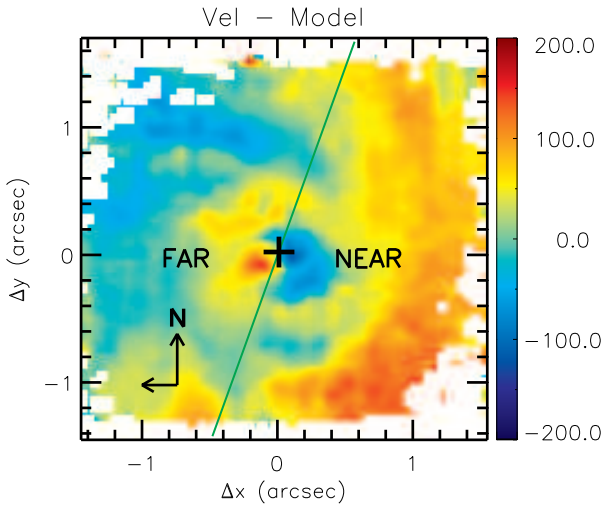}
\caption{\small Velocity residual map for H$_2\lambda 2.1218\mu$\,m obtained from the difference between the \H2 velocity field (Fig.~\ref{vels}) and the gas velocity model.
The map is in units of km\,s$^{-1}$.}
\label{flows}
\end{center}
\end{figure}

In order to isolate the non-circular components of the molecular gas velocity field, we fitted the H$_2$ velocity field using the Eq.~\ref{model-bertola}, but keeping fixed the geometric parameters of the disk and the systemic velocity derived from the fitting of the stellar velocity field. This procedure allows us to derive the maximal rotation, that is expected to be larger for the gas than for the stars, since the gas may is usually located in a thin disk (while the stars have a higher velocity dispersion).  The resulting best model is very similar to that for the stars (Fig.~\ref{model}), but with a velocity amplitude of $A=473\,\pm\,20\,$km\,$s^{-1}$. We have built a residual map between the \H2 velocity field (Fig.~\ref{vels})
and the fitted best disk model. This \H2 residual map is shown in Fig.~\ref{flows}. Assuming that the spiral arms are trailing, we conclude that the far side
of the galaxy is the North-East and the near side is the South-West.

Beyond the inner 0\farcs5 (73\,pc) of the \H2 residual map we find what seems to be a spiral arm in blueshift to the North and North-East of the nucleus in the far side of the galaxy, that could be due to gas flowing in towards the nucleus. In the near side of the galaxy we find mostly redshifts, that can also be interpreted as due to inflows.

We thus interpret most of the residuals observed in Fig.~\ref{flows} as being originated
from gas in the plane of the galaxy flowing towards the nucleus. 
This interpretation is supported by the channel maps in \H2 shown in Fig.~\ref{channel-h2}: in the channel maps from -245 km\,s$^{-1}$ down to -65 km\,s$^{-1}$, the blueshifted arm can be observed in the far side of the galaxy plane, while in the channel maps from 116 km\,s$^{-1}$ to 356 km\,s$^{-1}$, the redshifted arm can be observed in the near side of the galaxy plane, supporting inflows towards the line of nodes where a structure resembling a nuclear bar is observed in the lower velocity channels (blueshifts and redshifts). This bar runs from the North-West to the South-East, and is actually a bit tilted relative to the line of nodes. 

Within the inner 0\farcs5 (73\,pc), the residual map of Fig.~\ref{flows} shows blueshifts in the near side of the galaxy plane and redshifts in the far side along PA $\approx\,90^{\circ}$ that we attribute to an \H2 nuclear outflow. 
A compact outflow, but for the ionized gas, has been reported in previous optical long-slit 
observations by \citet{delgado02}, \citet{rosario}, as well as by
our group in \citet{allan} using optical integral field spectroscopy.

\citet{allan} has found two kinematic components in disc rotation in NGC\,2110, that they have called the ``warm" (higher velocity dispersion) and ``cold" (lower velocity
dispersion) disks (their Figs. 11 and 12). Their residual velocity maps for the warm component shows positive values to the North-East of the nucleus and negative values to the South-West of it,
in a very compact structure (within the inner arcsec). They interpreted this component as a nuclear outflow and we conclude it could be identified with the compact outflow we see in \H2\ (Fig.~\ref{flows}).
Similarly, the velocity residual map for the cold component of \citet{allan} shows some excess redshifts to the South-West and blueshifts to the North-East that are similar to the residuals found in the \H2 kinematics,
that we thus identify with the inflows we have found in \H2.

\subsubsection{Feeding of the AGN}\label{feeding} 

Once we have concluded that the molecular gas kinematics suggest the presence of inflows towards the center, we can use the velocity field maps
to estimate the mass inflow rate as follows:

\begin{equation}
\dot{M}_{\rm{H_2}} = 2\,m_{\rm p}\,N_{\rm {H_2}}\,v\,\pi\,r^2\,n_{\rm arms},
\label{massinflow}
\end{equation}
where $m_{\rm p}$ is the proton mass, 
$v = v_{\rm {obs}}/{\rm sen}\,i$ is the velocity of the gas flowing towards the center
in the plane of the galaxy,
$v_{\rm obs} = 80\,{\rm\,km\,s}^{-1}$ is the observed velocity in
\H2 the residual map and
 $i = 42^{\circ}$ is the disk inclination relative to the plane
of the sky \citep{delgado02}.
This value of $i$  is very similar to that obtained by fitting the stellar kinematics ($i\approx\,43^{\circ}$).
We adopt $r = 0.3'' (\approx\,44\,$pc) 
as the radius of a circular cross section in the nuclear arm to the North/North-East of the nucleus at $1''$ from the nucleus.
$N_{\rm {H_{2}}}$ is the molecular gas density and 
$n_{\rm arms} = 2$ is the number of spiral arms. 
We have assumed that
the molecular gas is located in a disk with radius 
$r_{\rm d} = 1.5''\,(220\,$pc) corresponding to the NIFS FOV and a typical thickness of
$h = 30\,$pc \citep{hicks09}.

The value of the gas density was obtained from:
\begin{equation}
N_{\rm{H_2}} = \frac{M_{\rm H_2}}{ 2\,m_{\rm p}\,\pi\,r_{\rm d}^2\,h }.
\label{densinf}
\end{equation}
We find $N_{\rm H_2}\approx\,6.22\,\times\,10^{-3}\,{\rm cm}^{-3}$,
which results in an inflow rate of
$\approx\,4.55\,\times\,10^{-4}\,{\rm M_{\odot}\,yr^{-1}}$.
The inflow rate deriver above is small, and might be only a fraction of the total molecular gas flowing towards 
the center of the galaxy, as the warm molecular gas we observe in the near-IR may only be
the heated skin of a much larger reservoir of cold molecular gas, as we have discussed in Sec.~\ref{disc-mass}. 
Using the conversion factor for the cold/warm molecular gas obtained in Sec.~\ref{disc-mass}, we find that the molecular gas mass inflow rate could be higher than 10\,M$_\odot$\,yr$^{-1}$,


\subsubsection{Feedback of the AGN}\label{feedback} 

As we have also found outflows within the inner 0\farcs5 (73\,pc), we can also calculate their power.
These outflows have been identified as the 
blueshifts and redshifts  in the residual velocity map of the molecular hydrogen (Fig.~\ref{flows})
along PA $\approx\,120^{\circ}$ within the inner $0\farcs5$. 

The outflow observed in $H_2$ has a similar orientation to that of the compact outflow seen in 
ionized gas by \citet{allan}, who also concluded that the outflow is not oriented along the ionization cone. Outlfows not co-spatial with the ionization cone have been observed 
for other active galaxies. 

\citet{ngc5929} report an equatorial outflow detected using NIFS observations for the 
Seyfert 2 NGC\,5929 \citep[see also][]{ngc5929a}.  For NGC\,2110 we proposed that the outflow is a result from gas in the plane of the galaxy being pushed
by clouds outflowing from the nucleus. The same hypothesis was put forth by \citet{allan}, 
where the authors observed a compact outflow from the nucleus of NGC2110, suggesting a spherical geometry, using optical integral field spectroscopy.
In the case of \H2, most of the gas is concentrated at the disk of the galaxy,  and thus the interaction of the expanding spherical bubble is seen only at locations where the \H2 is present, forming the bipolar structure seen in the residual map.

On the basis of our and other's previous studies
of the morphology of the outflow (as its extent is of the order of the PSF FWHM), we assume a bi-conical structure with an opening angle of 60$^{\circ}$ with a radius $r\approx\,31$\,pc 
(consistent with the constraints posed by the flux distribution of the outflowing gas)
and use the following expression:

\begin{equation}
\dot{M}_{\rm{out}} = \frac{2\,m_{\rm p}\,N_{\rm H_2}\,v\,\pi\,r^2\,n}{\rm sen\,\theta},
\label{mout}
\end{equation}
\noindent where $v$ is the outflowing gas velocity,
$ n=2$ accounts for the two sides of the bi-cone and
$\theta$ is the angle that the central axis of the cone makes with the plane
of the sky.
The velocity can be obtained directly from Figure~\ref{flows} and is
$v=70\,{\rm\,km\,s}^{-1}$.
Using the estimated $\theta = 18^{\circ}$ by \citet{rosario}, we obtain
$\dot{M}_{\rm out} = 4.3\,\times\,10^{-4}\,$M$_{\odot}$\,yr$^{-1}$.
But again, as for the mass  inflow rate, the total mass outflow rate, that is probably dominated by the cold molecular gas, can be 5--7 orders of magnitudes higher.
The mass outflow rate obtained for the ionized gas by \citet{allan}, is 0.9\,M$_\odot$\,yr$^{-1}$, thus much higher than that in the warm molecular gas, although it may be comparable to that in the cold molecular gas in NGC\,2110, as discussed above. This result for NGC\,2110 differs from those we have obtained for most galaxies we have studied so far, where the \H2 gas is usually rotating in the galaxy plane or in inflow. It is seldom found in outflow. Outflows are more frequently found in the ionized gas \citep[e.g.][]{allan,mrk1066c,sb10,mrk79}.

\citet{durre14} report the presence of a nuclear bar seen in the \feii flux map, but the velocity field observed in \feii\, shows only blueshifts, although this depends on the adopted systemic velocity. 
The position angle of this bar, as well as its extent ($\approx$90\,pc) is consistent with what we have interpreted as an outflow seen in our \H2 velocity map,
that shows blueshifts to one side of the nucleus but redshifts to the other side.
If this structure is a bar, one possibility is that the inflows are feeding the bar, whose orientation (P.A.$\sim$ 125$^{\circ}$) is close to the galaxy major axis (P.A.$\sim\,162^{\circ}$).
 Unfortunately we do not have observations in the J or H band to compare the
\feii emitting gas kinematics with that of the \H2 emitting gas. Further observations and modeling would be necessary in order to settle on the nature of this feature.


\subsubsection{AGN accretion rate}

The inflow and outflow rates estimated above can be
compared to the accretion rate onto the SMBH that is needed to power the AGN.
The mass accretion rate to the AGN can be estimated via:
\begin{equation}
\dot{m} = \frac{L_{\rm bol}}{c^2\eta},
\label{txacre}
\end{equation}
where $c$ is the speed of light,
$\eta$ is the efficiency of conversion of matter into energy (for
Seyfert galaxies $\eta \approx\,0.1$),
and $L_{\rm bol}$ is the AGN bolometric luminosity. 
The X-ray luminosity of the AGN in NGC\,2110 is estimated as
$L_X = 2.9\,\times\,10^{42}\,$erg\,s$^{-1}$ \citep{pellegrini}, 
and in order to obtain the bolometric luminosity we adopt the usual relation $L_{\rm bol} = 10\,L_X$.
The resulting accretion rate is thus: 
$\dot{m} = 1.6\,\times\,10^{-3}\,$M$_{\odot}$\,yr$^{-1}$.

If we use instead the bolometric luminosity from \citet{rosario}, of $L_{bol}=5\times10^{43}\,$erg\,s$^{-1}$,
we obtain a somewhat larger value of $\dot{m} = 2.8\,\times\,10^{-3}\,$M$_{\odot}$\,yr$^{-1}$ than
that obtained above.

The accretion rate is approximately one order of magnitude higher than the inflow and outflow rates in warm molecular gas, indicating that it is necessary more mass flowing towards the center to feed the SMBH, but, as pointed above, the inflowing mass is probably dominated by cold molecular gas.

The mass inflow rate in warm \H2 obtained for NGC\,2110 is similar to those obtained for other Seyfert galaxies \citep{sb10,n7582,mrk1066a} with similar luminosities to that of NGC\,2110. The outflow rate in warm molecular gas is about three orders of magnitude smaller than mass outflow rates we have observed in ionized gas for other
Seyfert galaxies \citep[e.g.][]{n4051,mrk79,sanchez09} and up to 6 orders of magnitude smaller than those observed for high luminosity Seyfert 2 galaxies
\citep[e.g.][]{mcelroy14}. But the compact outflow we have observed in warm \H2 for NGC\,2110 may represent only a small fraction of the total outflow from the nucleus of this galaxy, that is probably also dominated by cold molecular gas.
A possible scenario for the inner region of NGC\,2110 is that the inflows observed in \H2 
feed the central region of the galaxy and the accreted gas then triggers both star formation and the nuclear activity. Indeed, young stellar clusters have been observed around the nucleus of NGC2110 by \citet{durre14}.


\section{Conclusions}\label{conclusions}

Using integral field spectroscopy in the near-infrared obtained with the Gemini instrument NIFS, we have mapped the stellar and gaseous kinematics within the inner 
$\approx$\,250\,pc of the Seyfert 2 galaxy NGC\,2110, as well as the
ionized and molecular gas flux distributions at an angular
resolution $0.15''$. Our main conclusions are:

\begin{itemize}
\item The stellar kinematics shows velocity dispersion reaching $\approx$\,250\,km\,$s^{-1}$
around the nucleus as well as a rotation pattern with a similar velocity amplitude; 

\item From the relation $M_{\bullet} - \sigma_{\star}$, we estimate the mass of the
SMBH as $M_{\bullet} = 2.7\,^{+3.5}_{-2.1}\,\times\,10^{8}$\,M$_{\odot}$.

\item The flux distributions of the \H2 and Br$\gamma$
emission lines are different: while the Br$\gamma$ flux seems to be collimated along
PA $\approx\,-30^\circ$ (from South-East to North-West), the \H2 emission is observed over the whole FOV:

\item From the emission-line ratios, we conclude that the \H2 emission is due to thermal processes (heating X-rays from
the AGN and/or shocks), with an excitation temperature in the range $\approx$\,2100-2700\,K.

\item The warm molecular gas and ionized gas masses within the observed FOV were estimated as
$M_{\rm H_2}\,\approx\,1.4\,\times\,10^3\,{\rm M_{\odot}}$ and 
$M_{\rm H\,II}\,\approx\,1.7\,\times\,10^6\,{\rm M_{\odot}}$, respectively;

\item The velocity field of the gas shows a rotation pattern similar to that observed for the
stars. However, deviations from this pattern are observed, due to both inflows and outflows, more 
clearly observed in the molecular gas emission, as the ionized gas emission covers just a small part of the FOV;

\item Inflows in warm \H2 along spiral arms are observed beyond the inner $\approx$\,70\,pc, 
at a mass inflow rate of ${\rm 4.6\,\times\,10^{-4}\,M_{\odot}\,yr^{-1}}$;

\item Outflows in warm \H2 are observed within the inner $\approx$\,70\,pc, at a mass outflow rate of
${\rm 4.3\,\times\,10^{-4}\,M_{\odot}\,yr^{-1}}$.

The uncertainties in the above mass flow rates might be high due to assumptions about the gas density and geometry. 
They are one order of magnitude lower than the accretion rate to the AGN. But we point out that these rates are in warm molecular gas, that may be only the heated skin of a much larger cold molecular gas reservoir. 
If we consider that previous observations of both warm and cold molecular gas in nearby AGNs reveal that typical ratios between
the cold and warm molecular gas masses are in the range 10$^5-10^7$, the above flow rates may be as high
as a few to tens of solar masses per year. Similarly, the total mass of molecular gas may reach 10$^9$\MS.

\end{itemize}

\section*{Acknowledgements}

We acknowledge the referee for valuable suggestions that helped to improve the paper. This work is based on observations obtained at the Gemini Observatory, 
which is operated by the Association of Universities for Research in Astronomy,
Inc., under a cooperative agreement with the 
NSF on behalf of the Gemini partnership: the National Science Foundation (United States), the Science and Technology 
Facilities Council (United Kingdom), the National Research Council (Canada), CONICYT (Chile), the Australian Research 
Council (Australia), Minist\'erio da Ci\^encia e Tecnologia (Brazil) and South-EastCYT (Argentina).  
This research has made use of the NASA/IPAC Extragalactic Database (NED) which is operated by the Jet
 Propulsion Laboratory, California Institute of  Technology, under contract with the National Aeronautics and Space Administration.
This work has been partially supported by the Brazilian institution CNPq.
{\it R.A.R.} acknowledges support from FAPERGS (project N0. 12/1209-6) and CNPq (project N0. 470090/2013-8).

\label{lastpage}

\end{document}